\documentclass[10pt,conference]{IEEEtran}
\newif\ifarxivversion
\arxivversiontrue 
\ifarxivversion
\IEEEoverridecommandlockouts
\fi

\usepackage{cite}
\usepackage{amsmath,amssymb,amsfonts}
\usepackage{booktabs} 
\usepackage{graphicx}
\usepackage{listings}
\usepackage{makecell}
\usepackage{textcomp}
\usepackage{tikz}
\usepackage{xcolor}

\usepackage{soul}

\usepackage{algorithm}
\usepackage{algpseudocode}

\usepackage{forest}

\usepackage{hyperref}

\usepackage{quantumcircuit}
\usepackage{quantumcircuit_as}

\usetikzlibrary{calc}
\usetikzlibrary{arrows.meta, positioning, shapes.geometric} 

\usepackage[most]{tcolorbox}
\newtcolorbox{observationbox}[1]{
    colback=black!2,
    colframe=black!35,
    boxrule=0.5pt,
    arc=2pt,
    left=4pt,right=4pt,top=4pt,bottom=4pt,
    title={\textbf{Observation #1}},
    fonttitle=\small,
    coltitle=black
}

\usepackage{amsthm}

\tikzset{
    myarrow/.style={->, thick, red},
}
\pgfmathsetmacro{\gateHeight}{-3} 
\pgfmathsetmacro{\gateWidth}{1.5} 
\newcommand{\plotwidth}{0.85\linewidth}
\newcommand{\figurewidth}{0.85\linewidth}

\definecolor{codegreen}{rgb}{0,0.6,0}
\definecolor{codegray}{rgb}{0.5,0.5,0.5}
\definecolor{codepurple}{rgb}{0.58,0,0.82}
\definecolor{backcolour}{rgb}{0.98,0.98,0.98}
\definecolor{section-color}{RGB}{200, 30, 30}

\lstdefinestyle{cppstyle}{
    language=C++,
    backgroundcolor=\color{backcolour},   
    commentstyle=\color{codegreen},
    keywordstyle=\color{magenta},
    numberstyle=\tiny\color{codegray},
    stringstyle=\color{codepurple},
    basicstyle=\ttfamily\scriptsize, 
    numbers=left,
    breakatwhitespace=false,         
    breaklines=true,                 
    captionpos=b,                    
    keepspaces=true,                 
    numbersep=5pt,                  
    showspaces=false,                
    showstringspaces=false,
    showtabs=false,                  
    tabsize=2
}

\lstdefinestyle{cppstyleV2}{
  language=C++,
  basicstyle=\ttfamily\footnotesize,
  backgroundcolor=\color{backcolour},
  frame=single,
  rulecolor=\color{codegray!60},
  framerule=0.4pt,
  framesep=6pt,
  numbers=left,
  numberstyle=\tiny\color{codegray},
  stepnumber=1,
  numbersep=8pt,
  xleftmargin=2.2em,
  framexleftmargin=2.0em,
  tabsize=2,
  showspaces=false,
  showstringspaces=false,
  showtabs=false,
  keepspaces=true,
  columns=fullflexible,
  breaklines=true,
  breakatwhitespace=true,
  keywordstyle=\color{codepurple}\bfseries,
  commentstyle=\color{codegreen}\itshape,
  stringstyle=\color{magenta},
  aboveskip=0.8\baselineskip,
  belowskip=0.6\baselineskip,
  mathescape=true,
  captionpos=b
}

\def\BibTeX{{\rm B\kern-.05em{\sc i\kern-.025em b}\kern-.08em
    T\kern-.1667em\lower.7ex\hbox{E}\kern-.125emX}}
\begin{document}

\providecommand{\Description}[1]{}

\title{Fewer Histories, Faster Paths:\\
Distributed Quantum Circuit Feynman Simulation via History Reduction, Checkpointing, and Pruning%
\ifarxivversion\thanks{\textcopyright~2026 IEEE. Personal use of this material is permitted. Permission from IEEE must be obtained for all other uses, in any current or future media, including reprinting/republishing this material for advertising or promotional purposes, creating new collective works, for resale or redistribution to servers or lists, or reuse of any copyrighted component of this work in other works.}\fi}
\author{\IEEEauthorblockN{Frej Larssen, Luca Pennati, Erik M. Åsgrim, Ivy Peng, Stefano Markidis}
\IEEEauthorblockA{\textit{KTH Royal Institute of Technology}, Sweden\\
\{flarssen, pennati, erima, bopeng, markidis\}@kth.se}
}

\maketitle

\begin{abstract}
We present a distributed method for exact sparse-output quantum circuit simulation based on the pure Feynman sum-over-histories formulation. The method computes selected computational-basis amplitudes exactly and addresses the exponential growth of the path sum through a reduced history formulation based on internal-wire assignments, determinism propagation, artificial sources, pruning, and checkpointed reuse. Boundary constraints are propagated through deterministic and wire-preserving gates, and explicit branching variables are introduced only where residual ambiguity remains. Shared work across related histories is captured via an autotuned checkpointed partition. The parallel execution model combines decomposition over requested outputs with concurrent history evaluation, while a dynamic server-worker architecture mitigates load imbalance from irregular branching and pruning.

Across the circuit families studied, the method adapts to different structural regimes of the reduced history space: zero artificial sources for QFT under backward analysis, substantial speedups from checkpointing and autotuning for amplitude amplification, and a runtime-fidelity tradeoff from threshold pruning for QAOA. On quantum walk circuits, it reconstructs exact selected-output distributions up to 100 qubits and achieves 85\% parallel efficiency on 8,192 CPU cores of a supercomputer.
\end{abstract}


\section{Introduction}
Classical quantum circuit simulation remains a fundamental tool for analyzing quantum algorithms, validating circuit transformations, and computing exact properties of circuit outputs~\cite{NielsenChuang2010}. The dominant exact approach is state-vector simulation, which explicitly stores and updates the full computational state of a multiqubit circuit~\cite{haner20175,Smelyanskiy2016qHiPSTER,Jones2019QuEST,Haner2018ProjectQ}. This formulation gives complete access to all output amplitudes, but its memory cost grows exponentially with the number of qubits. Tensor-network methods and hybrid Schrödinger-Feynman methods reduce this cost in some regimes, but their performance depends strongly on circuit structure, partitioning choices, and contraction strategy~\cite{MarkovShi2008,Orus2014,gray2018quimb,Arute2019Supremacy,markov_quantum_2018}. The cost of stabilizer-based simulators and Pauli propagation scales mainly with the non-Clifford resources in the circuit, by operating on generators of stabilizers and propagating Pauli observables in the Heisenberg picture, respectively~\cite{bennink_unbiased_2017, Bravyi2019simulationofquantum,Rall2019PauliPropagation,rudolph_pauli_2025}. Pure Feynman simulation, the topic of this paper, occupies a different point in this simulation design space. It computes output amplitudes directly as sums over classical histories induced by the circuit~\cite{Feynman1982Simulating,RudiakGould2006PathSum,ferreira2023feynman}.

In many quantum circuit workloads, the relevant output is not the full final state vector but a selected set of computational-basis amplitudes. We therefore consider the following problem: given a circuit, a sparse input state, and a specified set of output bitstrings, compute exactly only those requested amplitudes. This setting arises naturally in circuit families such as amplitude amplification, phase estimation, quantum approximate optimization, and quantum walk. For such workloads, pure Feynman simulation is appealing because it evaluates requested outputs directly, but its cost is still governed by the exponential growth of the path sum with circuit branching~\cite{Feynman1982Simulating,RudiakGould2006PathSum,Boixo2018Graphical}.

This paper develops an exact sparse-output Feynman simulator that exploits circuit structure to reduce the number of histories that must be evaluated.


The paper makes the following contributions:
\begin{itemize}
    \item We formulate exact sparse-output quantum circuit simulation as a pure Feynman path-sum computation.
    \item We introduce a history-reduction method based on internal-wire reasoning, determinism propagation, and artificial sources.
    \item We develop a checkpointed evaluation scheme with autotuned partitioning and pruning to reduce repeated work across histories.
    \item We show that the resulting method admits an efficient distributed-memory realization and enables exact selected-output simulation on supercomputers.
\end{itemize}

\section{Preliminaries}
\label{sec:preliminaries}
Given a quantum circuit $C = U^{(m)} \cdots U^{(1)}$, an input basis state $i^{\mathrm{in}}$, and an output basis state $i^{\mathrm{out}}$, a Feynman simulator computes the transition amplitude~\cite{NielsenChuang2010,RudiakGould2006PathSum}
\begin{equation}
    \alpha^{i^{\mathrm{in}}, i^{\mathrm{out}}}
    =
    \bra{i^{\mathrm{out}}}
    U^{(m)} \cdots U^{(1)}
    \ket{i^{\mathrm{in}}}.
    \label{eq:fixed-in-out}
\end{equation}
For a sparse input state $\ket{\psi}$, the amplitude of a requested output basis state is obtained by summing these transition amplitudes over the support of the input state~\cite{NielsenChuang2010},
\begin{equation}
    \braket{i^{\mathrm{out}}|\psi'}
    =
    \sum_{i^{\mathrm{in}}}
    \braket{i^{\mathrm{in}}|\psi}\,
    \alpha^{i^{\mathrm{in}}, i^{\mathrm{out}}},
    \qquad
    \ket{\psi'} = C\ket{\psi}.
    \label{eq:specific-output}
\end{equation}

A Feynman simulator evaluates a requested amplitude as a sum over \emph{histories}~\cite{Feynman1982Simulating,RudiakGould2006PathSum}. For a fixed input-output pair, a history is a sequence of intermediate computational-basis states
\begin{equation}
    \phi = \bigl(i^{\mathrm{in}}, i^{(1)}, \dots, i^{(m-1)}, i^{\mathrm{out}}\bigr),
\end{equation}
and its contribution is the product of the gate transition amplitudes along that path~\cite{RudiakGould2006PathSum},
\begin{equation}
    \alpha_{\phi}
    =
    \prod_{t=1}^{m}
    U^{(t)}_{i^{(t-1)},\,i^{(t)}}.
\end{equation}
The total transition amplitude is then
\begin{equation}
    \alpha^{i^{\mathrm{in}}, i^{\mathrm{out}}}
    =
    \sum_{\phi \in \Phi}
    \alpha_{\phi},
    \label{eq:sum-over-histories}
\end{equation}
where $\Phi$ is the set of all histories starting in $i^{\mathrm{in}}$ and ending in $i^{\mathrm{out}}$. This is the circuit-level sum-over-histories form of the Feynman path-sum picture~\cite{Feynman1982Simulating,RudiakGould2006PathSum}.

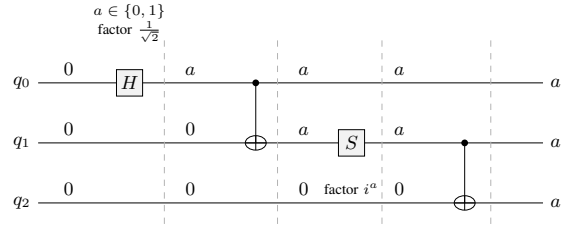
\begin{figure}[t]
\centering
\resizebox{\figurewidth}{!}{%
\begin{tikzpicture}[x=1.45cm,y=1.0cm, every node/.style={font=\small}]
    \draw (0,0) -- (5.8,0);
    \draw (0,-1) -- (5.8,-1);
    \draw (0,-2) -- (5.8,-2);

    \node[left] at (0,0) {$q_0$};
    \node[left] at (0,-1) {$q_1$};
    \node[left] at (0,-2) {$q_2$};

    \node[above] at (0.35,0) {$0$};
    \node[above] at (0.35,-1) {$0$};
    \node[above] at (0.35,-2) {$0$};

    \draw[fill=gray!10] (0.9,0.22) rectangle (1.2,-0.22);
    \node at (1.05,0) {$H$};
    \node[above, align=center] at (1.05,0.55) {\scriptsize $a\in\{0,1\}$\\[-1pt]\scriptsize factor $\frac{1}{\sqrt{2}}$};

    \node[above] at (1.75,0) {$a$};
    \node[above] at (1.75,-1) {$0$};
    \node[above] at (1.75,-2) {$0$};

    \fill (2.5,0) circle (1.6pt);
    \draw (2.5,0) -- (2.5,-1);
    \draw (2.5,-1) circle (0.12);
    \draw (2.38,-1) -- (2.62,-1);
    \draw (2.5,-0.88) -- (2.5,-1.12);

    \node[above] at (3.05,0) {$a$};
    \node[above] at (3.05,-1) {$a$};
    \node[above] at (3.05,-2) {$0$};

    \draw[fill=gray!10] (3.45,-0.78) rectangle (3.75,-1.22);
    \node at (3.6,-1) {$S$};
    \node[below, align=center] at (3.6,-1.55) {\scriptsize factor $i^a$};

    \node[above] at (4.15,0) {$a$};
    \node[above] at (4.15,-1) {$a$};
    \node[above] at (4.15,-2) {$0$};

    \fill (4.9,-1) circle (1.6pt);
    \draw (4.9,-1) -- (4.9,-2);
    \draw (4.9,-2) circle (0.12);
    \draw (4.78,-2) -- (5.02,-2);
    \draw (4.9,-1.88) -- (4.9,-2.12);

    \node[right] at (5.8,0) {$a$};
    \node[right] at (5.8,-1) {$a$};
    \node[right] at (5.8,-2) {$a$};

    \draw[dashed,gray!60] (1.45,-2.35) -- (1.45,0.75);
    \draw[dashed,gray!60] (2.75,-2.35) -- (2.75,0.75);
    \draw[dashed,gray!60] (3.95,-2.35) -- (3.95,0.75);
    \draw[dashed,gray!60] (5.2,-2.35) -- (5.2,0.75);
\end{tikzpicture}
}
\caption{Three-qubit example in the internal-wire formulation. The Hadamard creates the only branching variable, $a\in\{0,1\}$, with local amplitude factor $1/\sqrt{2}$. The $S$ gate preserves the wire value and contributes the phase factor $i^a$. The deterministic gates propagate the wire values so that the final output is $(a,a,a)$.}
\label{fig:three-qubit-wire-example}
\end{figure}

It is convenient to describe a Feynman history by the values carried by the wire segments between gates. Figure~\ref{fig:three-qubit-wire-example} illustrates the internal-wire formulation on a simple three-qubit circuit. Starting from the input state $\ket{000}$, the circuit applies a Hadamard gate on $q_0$, a CNOT from $q_0$ to $q_1$, an $S$ gate on $q_1$, and a final CNOT from $q_1$ to $q_2$. A Feynman history is described by the values carried by the wire segments between these gates.

The Hadamard on $q_0$ creates the only branching in the circuit. Let $a\in\{0,1\}$ denote the value of the top wire immediately after the Hadamard. This local transition contributes the amplitude
\[
\bra{a}H\ket{0}=\frac{1}{\sqrt{2}}
\qquad \text{for } a=0,1.
\]
All remaining wire values are then determined by the circuit structure. Since the first CNOT copies the control value from $q_0$ to $q_1$, the middle wire after that gate must equal $a$. The $S$ gate does not change the wire value on $q_1$, but contributes the phase factor
\[
S\ket{a}=i^a\ket{a}.
\]
Finally, the second CNOT copies the value on $q_1$ to $q_2$, so the output must be $(a,a,a)$. Therefore, only outputs of the form $(a,a,a)$ can receive nonzero amplitude, and
\begin{equation}
    \braket{b|C|000}
    =
    \begin{cases}
        \dfrac{1}{\sqrt{2}}, & b=(0,0,0),\\[6pt]
        \dfrac{i}{\sqrt{2}}, & b=(1,1,1),\\[6pt]
        0, & \text{otherwise.}
    \end{cases}
    \label{eq:three-qubit-wire-sum}
\end{equation}
There are therefore exactly two contributing histories, corresponding to $a=0$ and $a=1$, with amplitudes $1/\sqrt{2}$ and $i/\sqrt{2}$, respectively; all other output amplitudes are zero.

\subsection{Internal Wires and Gate Roles in the Feynman Simulator}
In the standard circuit model, a quantum circuit is an ordered sequence of gates acting on qubit wires from input to output~\cite{Mermin2007}. For a fixed computational-basis input and output, a Feynman history can be viewed as one assignment of intermediate basis values consistent with the circuit. As shown in Figure~\ref{fig:three-qubit-wire-example}, it is convenient to express that assignment in terms of the values carried by the circuit wires.

This leads to the \ul{internal-wire formulation} used in our simulator. The unknowns are the binary values on the wires between gates, excluding the input and output wires, which we call \emph{internal wires}. A naive formulation would treat every internal wire as an independent binary variable, but many of them can instead be inferred from gate constraints.

We therefore classify gates according to how they act on computational-basis wire values: whether they produce a unique output assignment, multiple possible output assignments, or the same assignment with only a phase change.
\begin{itemize}
    \item \textbf{Deterministic gates.}
    These are gates that map each computational-basis input configuration to a \emph{single} computational-basis output configuration on the wires they act on, possibly up to a phase factor. Once the input wire values are known, the output wire values are therefore fixed. This class includes basis-permuting gates such as $X$, $Y$, SWAP, CNOT, Toffoli, Fredkin, and, more generally, reversible classical logic embedded in a quantum circuit. For controlled permutation gates, the role is conditional: once the control values are known, the target action is either inactive (identity) or a fixed basis permutation, so no branching is introduced.

    \item \textbf{Branching gates.}
    These are gates for which a computational-basis input can produce \emph{multiple} computational-basis output configurations with nonzero amplitude. In the path-sum formulation, each such nonzero transition corresponds to a distinct continuation of the history. The canonical example is the Hadamard gate. More generally, any gate whose matrix has more than one nonzero entry in a column in the computational basis acts as a branching gate~\cite{markidis2024parallelism}. This includes, for example, generic single-qubit rotations such as $R_x(\theta)$ and $R_y(\theta)$ when both sine and cosine terms are nonzero. 

    \item \textbf{Wire-preserving gates.}
    These are gates that leave the computational-basis value on each acted-on wire unchanged and contribute only a multiplicative factor to the amplitude. In practice, this means that they are diagonal in the computational basis, or that they correspond to control wires whose values propagate unchanged while determining whether another operation is active. Examples include $Z$, $S$, $T$, $R_z(\theta)$, arbitrary diagonal phase gates, controlled-phase gates, and CZ. We distinguish this class from general deterministic gates because wire-preserving gates do not introduce new binary wire values: the value on each wire is identical on both sides of the gate, so the gate affects only the accumulated amplitude or the activation condition of a controlled operation.
\end{itemize}

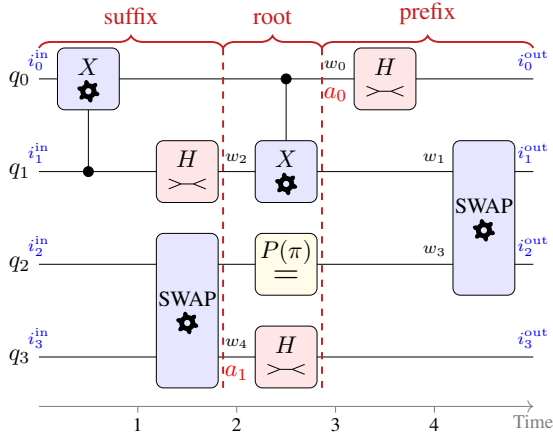
\begin{figure}[t]
    \centering
    \resizebox{\figurewidth}{!}{%
    \begin{tikzpicture}[ccircuit scale]
        \cdrawqubits{4}{5}

        \cgate{1}{0}{1}{$X$}          {0}{1}{0}
        \cgate{2}{1}{} {$H$}          {0}{0}{1}
        \cgate{2}{2,3}{} {\small SWAP}{0}{1}{0}
        \cgate{3}{3}{} {$H$}          {0}{0}{1}
        \cgate{3}{2}{} {$P(\pi)$}     {1}{0}{0}
        \cgate{3}{1}{0}{$X$}          {0}{1}{0}
        \cgate{4}{0}{} {$H$}          {0}{0}{1}
        \cgate{5}{1,2}{}{\small SWAP} {0}{1}{0}

        \ctimeaxis{5}{\ctimeunit}

        \cwlabel{1}{0}{${\scriptstyle \color{blue}{i^{\mathrm{in}}_0}}$}{}
        \cwlabel{1}{1}{${\scriptstyle \color{blue}{i^{\mathrm{in}}_1}}$}{}
        \cwlabel{1}{2}{${\scriptstyle \color{blue}{i^{\mathrm{in}}_2}}$}{}
        \cwlabel{1}{3}{${\scriptstyle \color{blue}{i^{\mathrm{in}}_3}}$}{}

        \cwlabel{4}{0}{${\scriptstyle w_0}$}{$\color{red}{a_0}$}
        \cwlabel{5}{1}{${\scriptstyle w_1}$}{}
        \cwlabel{3}{1}{${\scriptstyle w_2}$}{}
        \cwlabel{5}{2}{${\scriptstyle w_3}$}{}
        \cwlabel{3}{3}{${\scriptstyle w_4}$}{$\color{red}{a_1}$}

        \cwlabel{6}{0}{${\scriptstyle \color{blue}{i^{\mathrm{out}}_0}}$}{}
        \cwlabel{6}{1}{${\scriptstyle \color{blue}{i^{\mathrm{out}}_1}}$}{}
        \cwlabel{6}{2}{${\scriptstyle \color{blue}{i^{\mathrm{out}}_2}}$}{}
        \cwlabel{6}{3}{${\scriptstyle \color{blue}{i^{\mathrm{out}}_3}}$}{}

        \cdepthsections{5}{2}{1}{4}{}%
    \end{tikzpicture}
    }
    \caption{Internal-wire view of a four-qubit circuit. The figure shows a standard quantum circuit reinterpreted in terms of the binary values carried by its internal wires. Deterministic gates propagate wire values, branching gates introduce residual ambiguity, and wire-preserving gates leave wire values unchanged while contributing only phase or activation constraints. Blue labels indicate wire values fixed by the input and output bitstrings; red labels indicate residual branching variables, later formalized as artificial sources. The division into prefix, root and suffix represents checkpointing, an optimization later discussed.}
    \label{fig:circuit-internal-wires}
\end{figure}
Figure~\ref{fig:circuit-internal-wires} illustrates this simulator-oriented interpretation on a four-qubit circuit. The blue labels denote wire values fixed directly by the chosen input and output bitstrings, which form the initial boundary constraints and later serve as \emph{natural sources of determinism}. Deterministic gates propagate these values across the circuit, while wire-preserving gates leave them unchanged and contribute only phase or activation information. The remaining ambiguity is represented by the red labels $a_0$ and $a_1$, which mark the unresolved branching points that remain after this propagation.

\section{Related Work}
\textbf{Quantum Circuit Simulators.} Classical quantum circuit simulators differ mainly in how they represent the evolving quantum state and in the resulting tradeoffs among memory, computation, and communication. State-vector simulators, such as \texttt{Qiskit Aer} and \texttt{qsim}~\cite{Faj2023WarpSpeed,markidis2023enabling}, store the full $2^n$-dimensional state and apply gates directly to that representation. Distributed implementations such as \texttt{qHiPSTER}/\texttt{Intel-QS}~\cite{Smelyanskiy2016qHiPSTER}, \texttt{QuEST}~\cite{Jones2019QuEST}, and \texttt{ProjectQ}~\cite{Haner2018ProjectQ} extend this approach to larger systems, but remain limited by exponential memory growth and communication overhead. Related HPC simulator platforms include \texttt{qFlex}, developed for verification and benchmarking of large quantum circuits, and \texttt{HybridQ}, a flexible framework for integrating multiple simulation techniques across hardware backends~\cite{Villalonga2018qFlex,Mandra2021HybridQ}. Tensor-network methods reduce memory by expressing the circuit as a contracted network whose cost depends on topology, effective tree width, and contraction order~\cite{MarkovShi2008,Orus2014,Gray2021Hyper,gray2018quimb}. Their practical performance depends strongly on contraction planning and slicing. \\

\textbf{Feynman and Hybrid Simulators.} Hybrid Schr\"odinger-Feynman methods partition the circuit and sum over Schmidt terms across the cut, and have been central to simulations of quantum-supremacy-style circuits~\cite{Arute2019Supremacy,markov_quantum_2018,burgholzer_hybrid_2021}. Related compression-based ideas also appear in decision-diagram simulators~\cite{Viamontes2007QMDD}. In this work, we focus on \emph{pure Feynman simulation}, where amplitudes are evaluated directly as sums over classical circuit histories. The path-sum approach goes back to Feynman~\cite{Feynman1982Simulating}, with a circuit-level derivation in Ref.~\cite{RudiakGould2006PathSum}. Serial Feynman simulator implementations are developed in Ref.~\cite{ferreira2023feynman}, and related ideas have also been combined with graphical-model and decomposition techniques~\cite{Boixo2018Graphical,Bravyi2019simulationofquantum}. Closely related to the Feynman approach are Pauli-propagation and Pauli-path methods, which, like pure Feynman simulation, exploit path-sum structure to reduce computation, but do so in the Pauli basis and usually target observables or noisy-circuit expectation values rather than selected computational-basis amplitudes~\cite{Rall2019PauliPropagation,Aharonov2023NoisyRCS}.


\section{Method}
\label{sec:method}
Given a circuit $C$, a sparse input state $\ket{\psi}$, and a specified output set, the method evaluates each required transition amplitude using a pure Feynman path sum. The central goal is to avoid enumeration over all internal-wire assignments and instead sum only over the residual branching choices left after structural inference.

\subsection{Reduced-History Construction}
\label{subsec:reducedhistory}
Let $\mathcal{W}$ denote the set of internal wires of the circuit. A naive path-sum formulation would sum over all binary assignments $x \in \{0,1\}^{|\mathcal{W}|}$, most of which are either invalid or unnecessary. Our goal is to replace this with a much smaller sum over only the residual branching variables that remain after structural inference.

\noindent \textbf{Observation 1.} \textit{In the internal-wire formulation, only unresolved branching points contribute to the exponential growth of the history space. Deterministic and wire-preserving gates do not create new binary variables: they only propagate constraints on wire values or modify the accumulated amplitude.}

The simulator therefore avoids enumeration over full intermediate basis states. Instead, it maintains a partial internal-wire assignment and extends it only where the circuit admits multiple locally consistent continuations. Deterministic and wire-preserving implications are absorbed into a propagation step that infers additional wire values without increasing the number of candidate histories.

\noindent \textbf{Observation 2.} \textit{For a fixed input-output pair, the chosen input and output bitstrings already determine part of the internal-wire assignment, and in many circuits these boundary values determine additional internal wires through propagation. Backward propagation can delay explicit branching and further reduce the residual history space in most of the traditional quantum algorithms.}

We treat the wire values imposed directly by the input and output bitstrings as \emph{natural sources of determinism}. These boundary values are propagated before any history enumeration begins. If the resulting constraints are incompatible, then no history can connect the chosen input and output, and the transition amplitude is zero.

The remaining ambiguity is represented by a smaller set of explicitly introduced binary variables, which we call \emph{artificial sources}. Each artificial source corresponds to a branching point whose outgoing wire value is not already fixed by the natural sources or by previously propagated implications. Let $\mathcal{A}$ denote the set of artificial sources introduced by the analysis pass, with $A = |\mathcal{A}|$. Each candidate history is then specified by one binary assignment $a \in \{0,1\}^{A}$ together with all wire values implied by propagation, so the transition amplitude is reduced from a sum over all internal-wire assignments to a sum over artificial-source assignments:
\begin{equation}
    \alpha^{b_{\mathrm{in}},b_{\mathrm{out}}}
    =
    \sum_{a \in \{0,1\}^{A}} \chi(a)\,\alpha_a,
    \label{eq:reduced-history-sum}
\end{equation}
where $\chi(a)\in\{0,1\}$ indicates whether the assignment is consistent with the propagated constraints, and $\alpha_a$ is the contribution of the resulting valid history. The key point is that the exponential factor is governed by $A$, not by $|\mathcal{W}|$.

In our implementation, \ul{artificial-source placement is performed from output to input}. This backward sweep is valid because, for a fixed input-output pair, a contributing history is defined by an internal-wire assignment that is consistent with the local transition amplitudes of all gates. These consistency constraints can be propagated from either boundary of the circuit. For deterministic gates, which act as basis permutations, the induced wire values can be propagated uniquely in both directions. For wire-preserving gates, the same binary wire value appears on both sides of the gate, so the constraint also propagates unchanged in either direction. Only at branching gates is the continuation non-unique. These are therefore precisely the locations where the analysis may need to introduce an artificial source.

\noindent \textbf{Observation 3.} \textit{Many candidate histories can be rejected before full evaluation, either because they are inconsistent with the propagated wire constraints or because their partial contribution is too small to justify further refinement.}

This motivates \ul{two forms of pruning} in the sparse-output kernel.
\begin{itemize}
    \item \emph{Exact pruning} removes invalid histories as soon as inconsistency is detected. A candidate history is discarded if propagation forces two different values onto the same wire. This could be due to input and output bitstrings that are not compatible with a certain history, or suboptimal placement of artificial sources. In these cases, the corresponding history contributes exactly zero amplitude.
    \item \emph{Threshold-based pruning within a history} optionally discards branches whose running partial amplitude falls below a user-defined threshold $t$. When $t=0$, the method remains exact and only invalid histories are pruned. For $t>0$, the method trades fidelity for additional cost reduction.
\end{itemize}

Algorithm~\ref{alg:base-kernel} summarizes the resulting sparse-output kernel for one fixed input-output pair. The kernel first propagates all wire values implied by the natural sources, and then enumerates only the assignments to the artificial sources that remain.
\begin{algorithm}[t]
\caption{Reduced-history sparse-output kernel for one input-output pair}
\label{alg:base-kernel}
\begin{algorithmic}[1]
\Require Circuit $C$, input bitstring $b_{\mathrm{in}}$, output bitstring $b_{\mathrm{out}}$, pruning threshold $t$
\Ensure Transition amplitude $\alpha^{b_{\mathrm{in}}, b_{\mathrm{out}}}$

\State $C_0 \gets$ \textsc{SetBoundaryWires}$(C,b_{\mathrm{in}},b_{\mathrm{out}})$
\If{\textsc{PropagateKnownValues}$(C_0)$ fails}
    \State \Return $0$
\EndIf

\State $\alpha \gets 0$
\For{each assignment $h$ to the artificial sources of $C_0$}
    \State $C_h \gets$ copy of $C_0$
    \If{\textsc{SetArtificialSources}$(C_h,h)$ fails}
        \State \textbf{continue}
    \EndIf
    \If{\textsc{PropagateKnownValues}$(C_h)$ fails}
        \State \textbf{continue}
    \EndIf
    \State $\alpha_h \gets$ \textsc{EvaluateHistory}$(C_h,t)$
    \State $\alpha \gets \alpha + \alpha_h$
\EndFor
\State \Return $\alpha$
\end{algorithmic}
\end{algorithm}
The routine \textsc{SetBoundaryWires} fixes the values on the input wires from $b_{\mathrm{in}}$ and on the output wires from $b_{\mathrm{out}}$. The routine \textsc{PropagateKnownValues} repeatedly applies all deterministic and wire-preserving implications until no additional wire value can be inferred; it fails if two constraints require different values on the same internal wire. The routine \textsc{SetArtificialSources} assigns the residual branching variables chosen by $h$ and fails if one of them contradicts an already fixed wire value. These three routines implement exact pruning of invalid histories.

If propagation succeeds, the resulting resolved wire assignment defines a valid propagated history. The routine \textsc{EvaluateHistory} traverses the circuit once and multiplies the local gate factors induced by the resolved wire values. When $t=0$, it evaluates the history exactly. When $t>0$, it may terminate early if the magnitude of the running partial amplitude falls below the pruning threshold. \\[0.6em]
\noindent \textbf{Computational complexity.} For a fixed input-output pair, let $G$ denote the number of gates in the circuit and let $A$ denote the number of artificial sources introduced by the analysis pass. The reduced-history kernel considers at most $2^A$ candidate histories, and each valid candidate requires at most one propagation/evaluation pass through the circuit, with cost linear to the gate count $G$. The worst-case cost is therefore $O(2^A G)$ per input-output pair in exact mode. For a sparse input state with support size $S$ and a requested output set of size $M$, the total worst-case cost is $O(S M 2^A G)$. The exponential factor is governed by the number of artificial sources rather than by the total number of internal wires. In practice, exact pruning reduces the number of valid histories below this worst-case bound, and threshold-based pruning can reduce it further.

This can be compared to state-vector simulators, which scale exponentially with the number of qubits~\cite{Smelyanskiy2016qHiPSTER,Jones2019QuEST,Haner2018ProjectQ}, and stabilizer-based simulators, whose cost is governed by the amount of non-Clifford structure or magic resources in the circuit~\cite{bennink_unbiased_2017,Bravyi2019simulationofquantum}. In circuits with many qubits and many non-Clifford gates, but only a few branching gates, Feynman simulators such as ours may have an advantage. Such advantage claims are less straightforward against tensor-network-based simulators~\cite{MarkovShi2008,Orus2014,Gray2021Hyper,gray2018quimb}. For tensor networks to become costly, entanglement is required between states. For this entanglement to exist we need population of multiple states, which in turn requires branching and a potentially costly Feynman simulation. This is the reason we choose tensor networks as our baseline in the last experiment.

\subsection{Artificial source placement for general circuits}

\begin{figure}[t]
    \centering
    \resizebox{\figurewidth}{!}{%
    \begin{tikzpicture}
        \begin{scope}[ccircuitas scale]
            \cdrawqubitsas{3}{8}

            \cgateas{1}{2}{} {$H$}          {0}{0}{1}
            \cgateas{2}{1}{2}{$X$}          {0}{1}{0}
            \cgateas{3}{0}{2}{$X$}          {0}{1}{0}

            \cgateas{4}{1}{2}{$X$}          {0}{1}{0}
            \cgateas{5}{0}{2}{$X$}          {0}{1}{0}

            \cgateas{6}{1}{0}{$X$}          {0}{1}{0}
            \cgateas{7}{2}{1}{$X$}          {0}{1}{0}

            \cgateas{8}{0}{} {$H$}          {0}{0}{1}
            \cgateas{8}{1}{} {$H$}          {0}{0}{1}

            \ctimeaxisas{8}{\ctimeunitas}

            \cwlabelas{1}{0}{${\scriptstyle \color{blue}{i^{\mathrm{in}}_0}}$}{}
            \cwlabelas{1}{1}{${\scriptstyle \color{blue}{i^{\mathrm{in}}_1}}$}{}
            \cwlabelas{1}{2}{${\scriptstyle \color{blue}{i^{\mathrm{in}}_2}}$}{}

            \cwlabelas{8}{0}{${\scriptstyle w_0}$}{}
            \cwlabelas{5}{0}{${\scriptstyle w_1}$}{}
            \cwlabelas{8}{1}{${\scriptstyle w_2}$}{}
            \cwlabelas{6}{1}{${\scriptstyle w_3}$}{}
            \cwlabelas{4}{1}{${\scriptstyle w_4}$}{}
            \cwlabelas{7}{2}{${\scriptstyle w_5}$}{}

            \cwlabelas{9}{0}{${\scriptstyle \color{blue}{~~~i^{\mathrm{out}}_0}}$}{}
            \cwlabelas{9}{1}{${\scriptstyle \color{blue}{~~~i^{\mathrm{out}}_1}}$}{}
            \cwlabelas{9}{2}{${\scriptstyle \color{blue}{~~~i^{\mathrm{out}}_2}}$}{}
        \end{scope}

        \begin{scope}[shift={(1.0cm,-4.2cm)}]
            \input{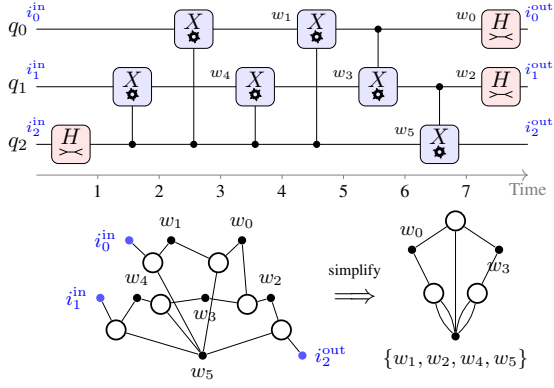}
        \end{scope}
    \end{tikzpicture}
    }
    \caption{Top: Circuit we want to find optimal artificial source placement of. Bottom: The corresponding graph with bullets representing sets of internal wires that are always reached together, and circles representing CNOTs.}
    \label{fig:as_placement}
\end{figure}

In this subsection, we briefly consider the problem of finding the minimal number of artificial sources needed for determinism to propagate, both forward and backward, to all internal wires in the circuit. We restrict our analysis to circuits consisting of the universal gate set CNOT + $R_x, R_y, R_z$ + phase shift $P$ + $H$. Single-qubit deterministic gates do not affect the artificial-source-placement solution, since determinism can propagate through them freely in either direction. We can therefore ignore them at this stage, and what remains are single-qubit branching gates, such as Hadamards, and CNOT.

An example is shown in Figure~\ref{fig:as_placement}. If the value of the control qubit of the CNOT, as well as one of the internal wires connected to the target qubit, is known, the other side of the target is also known. We also know that if both sides of the target are known, the control is given by xor-ing the target wires. From the viewpoint of determinism propagation, the three internal wires connected to a CNOT are therefore symmetric. This observation allows us to rewrite the circuit as a graph-like structure with three internal-wire nodes connected to each CNOT-node. If any two internal wires connected to a CNOT are reached by determinism, the third internal wire connected to that CNOT is also reached.
The problem of artificial source placement therefore reduces to finding the smallest set of nodes that must be initialized as reached so that all nodes in the equivalent graph eventually become reached.

A natural source connected to a CNOT (for example $i_1^{\mathrm{in}}$ in Figure~\ref{fig:as_placement}) means that the two other internal wires are reached at the same time ($w_4$ will be reached if and only if $w_5$ is). We can therefore simplify the graph form of the problem by applying a simple rule: for each natural source, remove the source, remove the CNOT it is connected to, and merge the two nodes.

Applying this rule to all natural sources yields the graph on the right side of the figure. In this simplified form, we only have three internal wires, compared to six in the original circuit. The number of ways to place artificial sources is therefore reduced. In the simplified graph of our example, we can conclude by brute force that the optimal solution is to place one artificial source on the merged node $\{w_1,w_2,w_4,w_5\}$; equivalently, placing an artificial source on any of these internal wires in the original circuit is enough to propagate values through the whole circuit.

A greedy algorithm could be implemented by iteratively 1) selecting the node in the simplified graph with the most edges and 2) simplifying the graph according to the same rule as for natural sources. Such an algorithm would aim to cover as much of the graph as possible with determinism at each step, but it would provide no guarantee of finding the optimal solution.

For the benchmark families in this paper, backward propagation attains the optimum, so we use it throughout.

\subsection{Checkpointing and Autotuned Partitioning}
\label{subsec:checkpointedparallel}

\noindent \textbf{Observation 4.} \textit{Many source assignments share the same propagated wire values and the same partial amplitude factors over contiguous circuit segments. The amount of reusable work therefore depends on where the circuit is partitioned.}

To exploit this structure, we partition the circuit into three contiguous regions and place checkpoints between them. Because histories are constructed by backward propagation, these regions are defined in the \emph{evaluation order} from output to input: the \emph{prefix} is the region adjacent to the output boundary, the \emph{root} is the middle region, and the \emph{suffix} is the remaining input-side region. Figure~\ref{fig:circuit-internal-wires} illustrates this organization on a four-qubit circuit. The regions are ordered according to the backward sweep from output to input: the prefix is adjacent to the output boundary, the root is the middle region, and the suffix is closest to the input boundary. In the example in Figure~\ref{fig:circuit-internal-wires}, the prefix contains one artificial source $a_0$, the root contains one artificial source $a_1$, and the suffix contains none. The checkpoint after the prefix is therefore reused across two root assignments, avoiding repeated evaluation of the prefix region.

Let $h_p$, $h_r$, and $h_s$ denote the assignments to the artificial sources introduced in the prefix, root, and suffix, respectively. For a fixed input-output pair, the backward sweep first processes the prefix. Each valid prefix assignment $h_p$ determines a propagated circuit state and a partial amplitude, denoted $(\sigma_p,\alpha_p)$, at the first checkpoint. All histories that share the same $h_p$ reuse this pair. The root region then refines that checkpoint: each valid $(h_p,h_r)$ produces a second propagated state and partial amplitude $(\sigma_r,\alpha_r)$, which are reused across all compatible suffix assignments. Finally, the suffix completes the history and contributes the remaining amplitude factor.

The benefit is visible in Figure~\ref{fig:circuit-internal-wires}. In that example, the prefix contains one artificial source $a_0$, the root contains one artificial source $a_1$, and the suffix introduces no new source, so $(A_p,A_r,A_s)=(1,1,0)$. Without checkpointing, the full circuit would be reevaluated for all four assignments $(a_0,a_1)\in\{0,1\}^2$. With checkpointing, the prefix is evaluated only once for each value of $a_0$, and each resulting checkpoint is then extended across the two possible values of $a_1$. Thus the work in the prefix is reused across multiple descendant histories instead of being recomputed from scratch.

\begin{algorithm}[t]
\caption{Checkpointed sparse-output kernel}
\label{alg:checkpointed-kernel}
\begin{algorithmic}[1]
\Require Partitioned circuit $C=(C_p,C_r,C_s)$, input bitstring $b_{\mathrm{in}}$, output bitstring $b_{\mathrm{out}}$
\Ensure Transition amplitude $\alpha^{b_{\mathrm{in}}, b_{\mathrm{out}}}$

\State $S_0 \gets$ \textsc{InitializeState}$(C,b_{\mathrm{in}},b_{\mathrm{out}})$
\If{\textsc{PropagateKnownValues}$(S_0)$ fails}
    \State \Return $0$
\EndIf

\State $\alpha \gets 0$
\For{each $h_p \in \{0,1\}^{A_p}$}
    \State $(S_p,\beta_p) \gets$ \textsc{AdvanceCheckpoint}$(S_0,C_p,h_p,1)$
    \If{$S_p=\bot$} \State \textbf{continue} \EndIf
    \For{each $h_r \in \{0,1\}^{A_r}$}
        \State $(S_r,\beta_r) \gets$ \textsc{AdvanceCheckpoint}$(S_p,C_r,h_r,\beta_p)$
        \If{$S_r=\bot$} \State \textbf{continue} \EndIf
        \For{each $h_s \in \{0,1\}^{A_s}$}
            \State $\beta_s \gets$ \textsc{AdvanceCheckpoint}$(S_r,C_s,h_s,\beta_r)$
            \If{$\beta_s\neq\bot$}
                \State $\alpha \gets \alpha + \beta_s$
            \EndIf
        \EndFor
    \EndFor
\EndFor
\State \Return $\alpha$
\end{algorithmic}
\end{algorithm}
Algorithm~\ref{alg:checkpointed-kernel} summarizes the resulting checkpointed kernel. Each region is processed by the same subroutine: assign the artificial sources introduced in that region, propagate all deterministic and wire-preserving implications within that region, and, if successful, multiply the corresponding local gate factors into the running amplitude. The propagated circuit state and partial amplitude at the checkpoint are then reused by all descendant branches. The routine \textsc{AdvanceCheckpoint} returns either failure, denoted by $\bot$, or a pair consisting of the propagated circuit state at the next checkpoint and the updated partial amplitude. \\[0.6em]
\textbf{Reduction of Gate Operations via Checkpointing.} Let $G_p$, $G_r$, and $G_s$ denote the numbers of gates in the prefix, root, and suffix, and let $A_p$, $A_r$, and $A_s$ denote the corresponding numbers of artificial sources. Ignoring pruning, the checkpointed kernel evaluates the prefix once for each of the $2^{A_p}$ prefix assignments, the root once for each pair of prefix and root assignments, and the suffix once for each complete assignment. This yields the gate-operation estimate
\begin{equation}
    Gate_{ops}
    =
    2^{A_p}
    \left(
        G_p + 2^{A_r}\left(G_r + 2^{A_s}G_s\right)
    \right),
    \label{eq:gateops}
\end{equation}
which makes the reuse pattern explicit. For the circuit in Figure~\ref{fig:circuit-internal-wires}, where $(A_p,A_r,A_s)=(1,1,0)$, the estimate becomes
\begin{equation}
    2\bigl(G_p + 2(G_r + G_s)\bigr).
\end{equation}
Evaluating all histories independently would require
\begin{equation}
    2^{A_p+A_r+A_s}(G_p+G_r+G_s)=4(G_p+G_r+G_s).
\end{equation}
Thus checkpointing reduces repeated work in the regions closer to the output boundary, where many descendant histories still share the same propagated state.\\[0.6em]
\textbf{Autotuned Partitioning.} Checkpoint placement is determined by an autotuning pass over candidate prefix/root/suffix splits. Each candidate defines two checkpoint locations and therefore a specific decomposition of the circuit into three regions. For every such decomposition, we construct the corresponding source placement, count the number of gates in each region, $(G_p,G_r,G_s)$, and count the artificial sources introduced in each region, $(A_p,A_r,A_s)$. These six quantities define the estimated cost of the checkpointed kernel through Equation~\eqref{eq:gateops}. The autotuner evaluates this estimate for all candidate decompositions and selects the partition with the smallest predicted gate-operation count. The resulting partition is therefore adapted to the circuit’s actual branching structure rather than chosen from depth alone.

\subsection{Parallel Decomposition and Execution Model}
\label{subsec:parallel-model}

\noindent \textbf{Observation 5.} \textit{For a sparse-output Feynman simulator, the amplitudes of different requested output bitstrings are independent.}

This independence provides a natural coarse-grained decomposition of the workload. Let $\mathcal{B}_{out}$ denote the requested set of output bitstrings, and let $\mathrm{supp}(\psi)$ denote the support of the sparse input state. For each $b_{\mathrm{out}}\in\mathcal{B}_{out}$, the simulator computes
\[
\braket{b_{\mathrm{out}}|\psi'}=
\sum_{b_{\mathrm{in}}\in\mathrm{supp}(\psi)}
\braket{b_{\mathrm{in}}|\psi}\,
\alpha^{b_{\mathrm{in}},b_{\mathrm{out}}},
\]
and no term associated with one output bitstring depends on the computation associated with another. The \ul{first level of parallel decomposition is performed over output bitstrings}. Each process is assigned a subset of $\mathcal{B}_{out}$ and evaluates the corresponding sparse-output kernels independently. This is a major advantage over state-vector and tensor-network simulation, where parallelization is often more challenging because the evolving state or contraction structure must be distributed and synchronized across processes.

\noindent \textbf{Observation 6.} \textit{Although output amplitudes are independent, their computational cost is highly irregular. Different output bitstrings can induce very different numbers of valid histories, different numbers of surviving checkpoint branches, and different amounts of pruning.}

This irregularity makes static output decomposition prone to load imbalance. A balanced execution model must therefore distribute work dynamically rather than assume that all output bitstrings have comparable cost. We address this with an \ul{asynchronous server-worker scheme}. One process acts as a server and maintains a global queue of output-bitstring batches, while the remaining processes act as workers. Each worker repeatedly requests a batch, evaluates the corresponding sparse-output kernels locally, and requests more work when it becomes idle. Within each worker, the kernel itself exposes a second level of parallelism. In the reduced-history kernel, the natural unit of concurrency is one assignment to the artificial sources. In the checkpointed kernel, the most effective parallelization level is typically the prefix level, since all descendant root and suffix refinements associated with a fixed prefix assignment share the same propagated prefix state and partial amplitude. 

\subsection{Implementation}
\label{subsec:implementation}
We realize this execution model in a distributed-memory implementation written in C++ and based on MPI and OpenMP, and made available as an open-source software \footnote{\url{https://github.com/frejlarssen/Feynman}}. MPI is used for the inter-process decomposition over output bitstrings \cite{Gropp1999MPI}, while OpenMP is used to exploit intra-kernel concurrency inside each worker \cite{Chapman2007OpenMP}.

\begin{figure}[t]
    \centering
    \resizebox{\figurewidth}{!}{
\begin{tikzpicture}[
    scale=0.6, every node/.style={transform shape},
    box/.style={rectangle, draw, fill=gray!5, rounded corners, minimum width=6cm, minimum height=4.5cm, align=center},
    innerbox/.style={rectangle, draw, rounded corners, minimum width=1cm, minimum height=1cm, align=center},
    innerrect/.style={rectangle, draw, fill=gray!20, text width=2.5cm, minimum height=1cm, align=center, font=\large},
    proc/.style={circle, draw, fill=blue!20, minimum size=0.3cm, align=center},
    file/.style={rectangle, draw, fill=orange!20, text width=2.1cm, minimum height=0.9cm, align=center},
    label/.style={font=\Large},
    innerlabel/.style={font=\large},
    singlearrow/.style={-{Stealth}, thick},
    doublearrow/.style={Stealth-Stealth, thick},
    node distance=1.0cm and 1.0cm
]

\node[box] (st1box) {};
\node[label, below=0cm of st1box.north] (st1label) {\textbf{ST.1: Initialization}};
\node[innerrect, below=0.1cm of st1label] (st1inner1) {Read input};
\node[innerrect, below=0.1cm of st1inner1] (st1inner2) {Build circuit};
\node[innerrect, below=0.1cm of st1inner2] (st1inner3) {Autotune};

\node[box, right=of st1box, inner sep=0.5cm] (st2box) {};

\node[label, below=0cm of st2box.north] (st2label) {\textbf{ST.2: Compute}};
\node[above=0.0cm of st2label.south, align=center, text width=4.5cm] (st2text) {%
};

\node[proc, fill=green!20, below=0.1cm of st2label] (st2server) {MPI\\Server};
\node[proc, below left=0.8cm and 0.8cm of st2server] (st2w1) {MPI\\W1};
\node[proc, below=0.8cm of st2server] (st2w2) {MPI\\W2};
\node[proc, below right=0.8cm and 0.8cm of st2server] (st2w3) {MPI\\W3};

\draw[doublearrow] (st2server) -- (st2w1);
\draw[doublearrow] (st2server) -- (st2w2);
\draw[doublearrow] (st2server) -- (st2w3);

\draw[singlearrow] (st1box.east) -- (st2box);

\node[box, below=of st2box, inner sep=0.5cm] (st3box) {};

\node[label, below=0cm of st3box.north] (st3label) {\textbf{ST.3: Write output}};

\node[proc, below=0.1cm of st3label] (w2st2) {MPI\\W2};
\node[file, below=1.1 of w2st2] (outfile) {Output file};
\node[proc, above left=0.8cm and 0.6cm of outfile] (w1st2) {MPI\\W1};
\node[proc, above right=0.8cm and 0.6cm of outfile] (w3st2) {MPI\\W3};

\draw[singlearrow] (w1st2) -- (outfile);
\draw[singlearrow] (w2st2) -- (outfile);
\draw[singlearrow] (w3st2) -- (outfile);

\draw[singlearrow] (st2box) -- (st3box);

\node[box, draw=blue, fill=blue!5, below left=of st2box, inner sep=0.5cm] (worker) {};
\node[label, below=0cm of worker.north] (workerlabel) {\textbf{Worker}};

\node[innerbox, fill=purple!15, below=0.4cm of workerlabel.center, text width=5.0cm, minimum height=3.5cm](kernel){};
\node[innerlabel, below=0cm of kernel.north] (kernellabel) {\textbf{Kernel}};

\node[innerbox, fill=red!25, below left=0.6cm and 0.2cm of kernellabel.center] (omp1) {OpenMP 1\\ \texttt{history 1}};
\node[innerbox, fill=red!25, below right=0.6cm and 0.2cm of kernellabel.center] (omp2) {OpenMP 2\\ \texttt{history 2}};
\node[innerbox, fill=red!25, below=0.4cm of omp1] (omp3) {OpenMP 3\\ \texttt{history 3}};


\node[innerbox, fill=red!25, below=0.4cm of omp2] (omp4) {OpenMP 4\\ \texttt{history 4}};

\draw[singlearrow, blue] (st2w1) -- (worker);

\end{tikzpicture}}
    \caption{Implementation workflow and two-level parallel execution of the sparse-output Feynman simulator. In ST.1, all MPI processes read the input and construct the circuit representation, including source analysis and autotuning. In ST.2, an MPI server distributes output batches dynamically to worker processes, which evaluate sparse-output kernels locally using OpenMP. In ST.3, workers write their local results in parallel to the output file.}
    \label{fig:software-organization}
\end{figure}
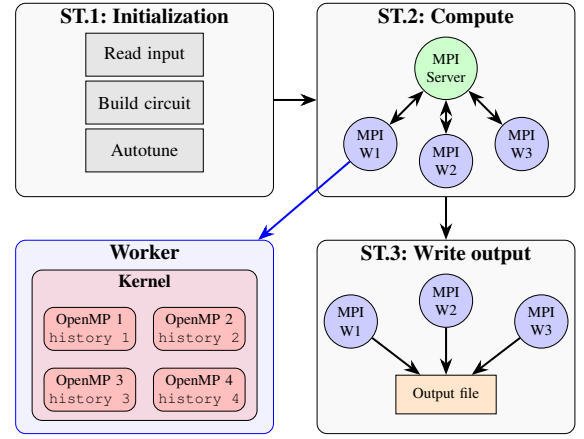

Figure~\ref{fig:software-organization} summarizes the implementation workflow. In \textbf{ST.1}, all MPI processes read the sparse input description, the requested output list, and the circuit, construct the internal circuit representation, and perform the preprocessing required by the method, including source placement and autotuning of the checkpoint partition. In \textbf{ST.2}, one MPI process assumes the role of server and maintains the global work queue over output batches. The worker processes repeatedly receive a batch of outputs, evaluate the associated sparse-output kernels locally, and return for more work when their current batch has been completed. In \textbf{ST.3}, each worker stores its local results and writes them in parallel to the output file.

The MPI layer is asynchronous. Workers request new batches only when they become idle, and the server responds dynamically according to the remaining workload. This design minimizes load imbalance caused by output-level irregularity and avoids the poor utilization that would arise from a static partitioning of the output list. Since all workers hold the circuit representation and the requested output list locally, the communication volume is small: the server needs only to communicate the identity and size of the next batch.

OpenMP is used inside each worker to evaluate the local sparse-output kernels concurrently. In the non-checkpointed kernel, each OpenMP task corresponds naturally to one artificial-source assignment. In the checkpointed kernel, each task is associated with one prefix assignment together with its descendant root and suffix refinements, so that the propagated checkpoint state and partial amplitude are reused within the task. This maps the software implementation directly onto the checkpointed execution model developed in the previous subsection.

\subsection{Circuits of Interest}

In the evaluation of the simulator we consider the Quantum Fourier Transform (QFT)~\cite{johnston2019programming}, Amplitude Amplification~\cite{johnston2019programming}, Quantum Approximate Optimization Algorithm (QAOA) for the max cut problem~\cite{farhi_quantum_2014}, and the quantum walk (QW)~\cite{kempe2003quantum}.

For the max cut problem in QAOA we choose a ring structure for the edges.

\begin{table}[!ht]
    \centering
    \caption{Depth and artificial sources across representative circuit families. $n$ is the number of qubits and $s$ is the number of steps/repetitions/iterations of Amplitude Amplification, QAOA, and QW.}
    \label{tab:algorithmic_metrics}
    \begin{tabular}{lcccc}
    \toprule
    Circuit & Depth & $A^{forward}$ & $A^{backward}$ \\
    \midrule
    QFT &  $O(n^2)$ & $n$ & $0$  \\
    Amplitude Amplification & $O(sn)$ & $2sn$ & $2sn$  \\
    QAOA (Max Cut) & $O(sE)$ & $sn$ & $sn$  \\
    QW &  $O(sn)$ & $s$ & $s-1$  \\
    \bottomrule
    \end{tabular}
\end{table}

By investigating the graphical representations of the circuits we can easily observe the scaling of the depth and the number of artificial sources. In Table~\ref{tab:algorithmic_metrics} we provide the scaling of these metrics for forward propagation and backward propagation of determinism. We observe that backward propagation results in fewer or equal number of artificial sources for all studied circuits.

\section{Results}
We evaluate the method through a sequence of application-driven case studies, each chosen to highlight a different aspect of the algorithm. We begin with an algorithmic view of the reduced-history formulation, focusing on how runtime depends on the number of artificial sources introduced by the circuit analysis. We then use QFT to validate exact sparse-output reconstruction in a regime with maximal history reduction, amplitude amplification to isolate the effect of checkpointing and autotuning, QAOA to study pruning, and quantum walk to demonstrate large-scale sparse-output reconstruction and distributed execution.

\begin{figure}[t]
    \centering
    \includegraphics[width=\plotwidth]{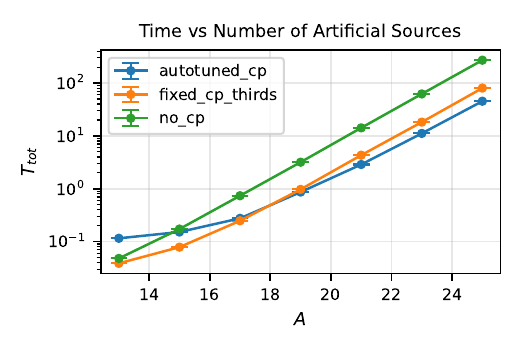}
    \caption{Time consumption versus number of artificial sources for quantum walk simulations. The dominant growth reflects the exponential dependence of the reduced-history kernel on the number of artificial sources, while the spread at fixed source count is caused by differences in pruning and checkpoint reuse.}
    \label{fig:artificial_sources_time_qw}
\end{figure}

The most important complexity variable of the method is the number of artificial sources introduced by the circuit analysis. For a fixed input-output pair, the reduced-history kernel considers at most $2^A$ candidate histories, where $A$ is the number of artificial sources, and this exponential dependence is the main determinant of runtime. Figure~\ref{fig:artificial_sources_time_qw} therefore examines the relationship between execution time and the number of artificial sources for quantum walk circuits.

The figure shows that the number of artificial sources is the primary driver of runtime, but not the only one. Outputs with the same value of $A$ can still differ in cost because exact pruning removes inconsistent branches, threshold pruning may truncate low-contribution branches, and checkpointing reuses partial work across related histories. \\

\begin{figure}[t]
    \centering
    \includegraphics[width=\plotwidth]{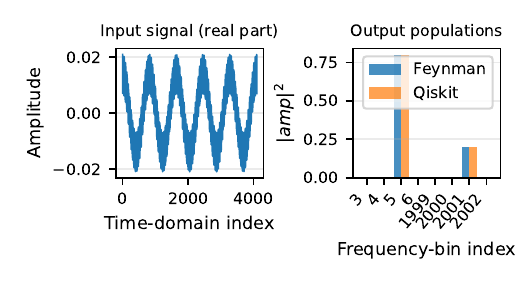}
    \caption{Validation of sparse-output amplitudes for a 12-qubit QFT circuit. Left: real part of the input signal in the time domain over 4096 basis states. Right: output populations for a selected set of requested frequency bins. Both our Feynman simulator and the \texttt{Qiskit Aer} simulator identify nonzero population only at bins 5 and 2,001, with matching amplitudes.}
    \label{fig:qft_validation}
\end{figure}

\noindent \textbf{Quantum Fourier Transform.}
QFT illustrates the most favorable algorithmic regime of the method. The backward propagation eliminates branching entirely, and sparse-output evaluation naturally targets only selected frequency bins. Figure~\ref{fig:qft_validation} validates the correctness of the method on Quantum Fourier Transform (QFT) circuits against \texttt{Qiskit Aer} state-vector simulator. This example also illustrates the sparse-output setting targeted by the method. In frequency analysis, one is often interested only in a selected subset of output bins rather than in the full transformed state.

From the algorithmic point of view, QFT is particularly well suited to backward history reduction. The controlled-phase structure on the output side is wire-preserving, and the Hadamard gates are adjacent to the input boundary. As a result, the backward sweep reaches the input side before encountering unresolved branching. No artificial sources are introduced, and each fixed input-output pair has at most one valid history. In this regime, the runtime is governed primarily by the number of requested output bins and by the sparsity of the input state, rather than by exponential branching internal to the circuit.

In Figure~\ref{fig:qft_validation}, the input is a 12-qubit real-valued signal in the time domain, and the sparse-output evaluation is restricted to a small set of candidate frequency bins around two expected peaks. The simulator identifies nonzero amplitudes only at bins 5 and 2,001, in agreement with \texttt{Qiskit Aer} simulator, while the remaining requested bins are zero. \\

\begin{figure}[t]
\centering
\includegraphics[width=\plotwidth]{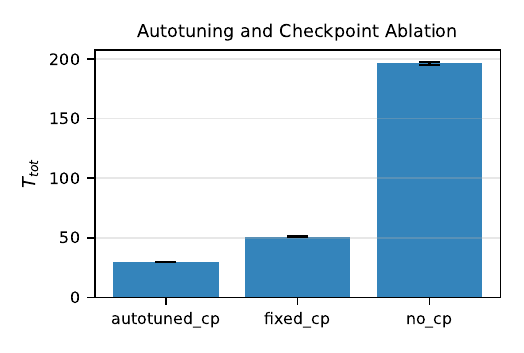}
\caption{Checkpointing ablation for amplitude amplification with 4 qubits and 3 iterations, $\ket{0}$ as input, and the lowest 10 computational-basis states requested as outputs. The bars compare no checkpointing, checkpointing with a fixed partition (a third of the gates in each region), and checkpointing with an autotuned partition.}
\label{fig:checkpointing_ablation}
\end{figure}

\noindent \textbf{Amplitude Amplification.}
Amplitude amplification provides a test case for evaluating the effect of checkpointing and autotuning on the reduced-history kernel. We consider a circuit with four qubits and three Amplitude Amplification iterations, chosen to be small enough to remove distributed-execution effects from the measurement while still exhibiting a nontrivial reduced-history space. Figure~\ref{fig:checkpointing_ablation} compares three execution modes: no checkpointing, checkpointing with a fixed partition, and checkpointing with an autotuned partition.

The figure shows that checkpointing substantially reduces the total execution time even for this small circuit. Relative to the baseline without checkpointing, a fixed partition lowers the runtime by roughly a factor of three, while autotuning leads to a further reduction and improves the runtime by about a factor of five overall. The gain comes from eliminating repeated evaluation of shared circuit segments. Without checkpointing, each source assignment traverses the full circuit independently. With checkpointing, histories that agree on earlier source assignments reuse the same propagated checkpoint state and partial amplitude.  \\

\noindent \textbf{QAOA.}
Threshold-based pruning is most informative in circuits whose branching histories carry strongly nonuniform weights, and QAOA provides a natural example of this regime. In QAOA, the $R_x$ layers act as branching gates, with different weights for different branches, while the phase-separation layers modulate the resulting history amplitudes by a phase factor. In this case, many branches contribute only weakly to the final amplitudes and become good candidates for early truncation.

\begin{figure}[t]
    \centering
    \includegraphics[width=\plotwidth]{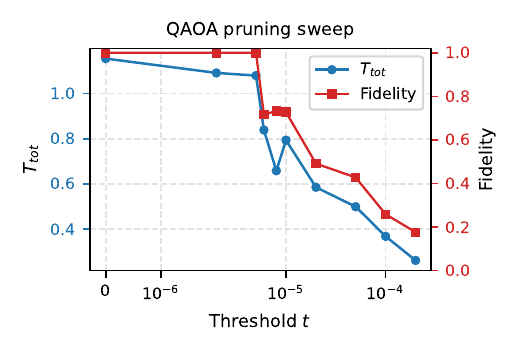}
    \caption{Pruning sweep for a QAOA instance. The threshold $t$ controls early truncation of low-contribution histories. As $t$ increases, the total execution time $T_{\mathrm{tot}}$ decreases, but only up to a point without significant loss in fidelity. Beyond a threshold near $10^{-5}$, further runtime reduction is accompanied by a rapid degradation in output fidelity.}
    \label{fig:qaoa_pruning_sweep}
\end{figure}

We sweep the pruning threshold $t$ and measure both total execution time and output fidelity. Figure~\ref{fig:qaoa_pruning_sweep} shows a clear tradeoff. For thresholds up to about $10^{-5}$, the fidelity remains close to one while the runtime decreases only moderately. For larger thresholds, the runtime drops more substantially, but the fidelity deteriorates rapidly. In the plotted instance, the total runtime is reduced by nearly an order of magnitude across the full sweep, while fidelity falls from essentially exact reconstruction to roughly one half. The useful operating regime is therefore the intermediate threshold range, where low-contribution histories are pruned aggressively enough to reduce cost, but not so aggressively that the output distribution is significantly distorted. \\


\noindent \textbf{Quantum Walk.}
Quantum walk circuits provide the main demonstration of the method, because they simultaneously expose sparse-output structure, irregular per-output cost, and large-scale parallelism. We consider a 100-qubit quantum walk in which the least significant qubit encodes the coin and the remaining qubits encode the position. The walker is initialized at position $2^{98}$ and the coin qubit in the state $(\ket{0}+i\ket{1})/\sqrt{2}$. We restrict the simulation to the 2{,}000 output bitstrings corresponding to 1{,}000 walker positions to the left and right of the initial position.

\begin{figure}[t]
    \centering
    \includegraphics[width=\plotwidth]{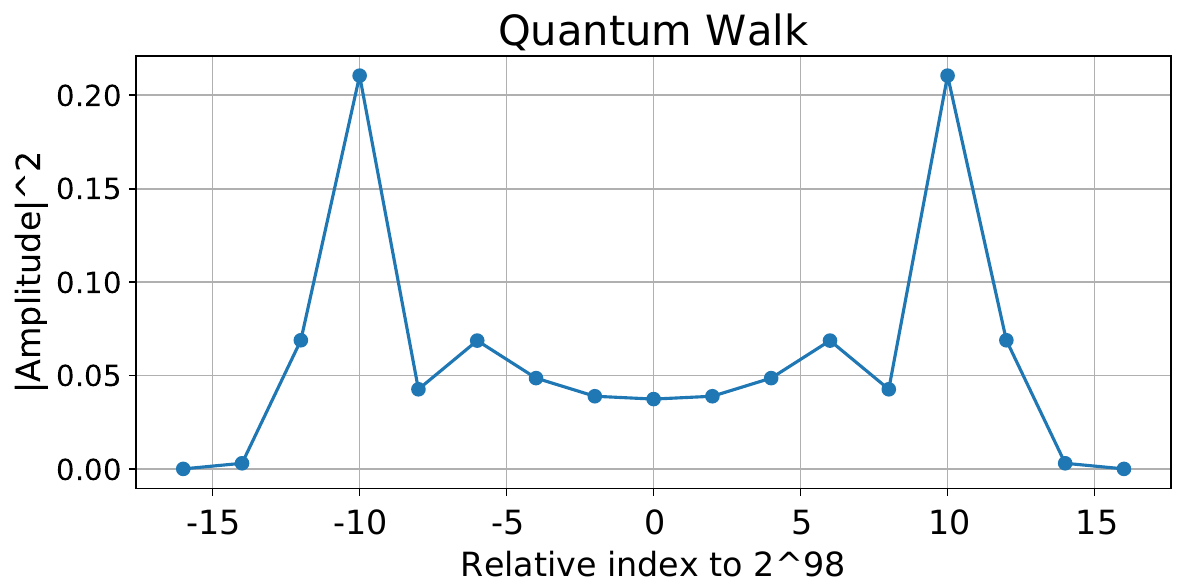}
    \caption{Exact sparse-output reconstruction for a 100-qubit quantum walk of 16 steps. Positions are reported relative to the initial position $2^{98}$. The coin qubit is initialized in the state $\frac{1}{\sqrt{2}}(\ket{0}+i\ket{1})$, yielding the characteristic symmetric probability distribution. The simulation was run with 16 MPI ranks on one node of System A in 73 seconds.}
    \label{fig_qrw100_super}
\end{figure}

Figure~\ref{fig_qrw100_super} shows the resulting probability distribution. The reconstructed output exhibits the expected symmetric profile, and the sum over the squared amplitudes is equal to one. In this experiment, the selected sparse-output set therefore captures all nonzero amplitudes of the final state, showing that complete observable reconstruction can be obtained without evaluating the full state vector.

The same circuit family is also used to evaluate the distributed-memory realization of the method. The scaling study is performed on two production systems with different processor and memory architectures: System~A, based on AMD EPYC 7742 CPUs with large DDR memory capacity per node, and System~B, based on Fujitsu A64FX CPUs with HBM2 memory. In both cases, the code is compiled with the \texttt{-O2} optimization flag, each MPI rank uses eight OpenMP threads, and ranks are placed so that threads are pinned to physical cores. For the experiments reported below, MPI ranks are distributed 16 per node on System~A and 6 per node on System~B.

\begin{figure}[t]
    \centering
    \includegraphics[width=\plotwidth]{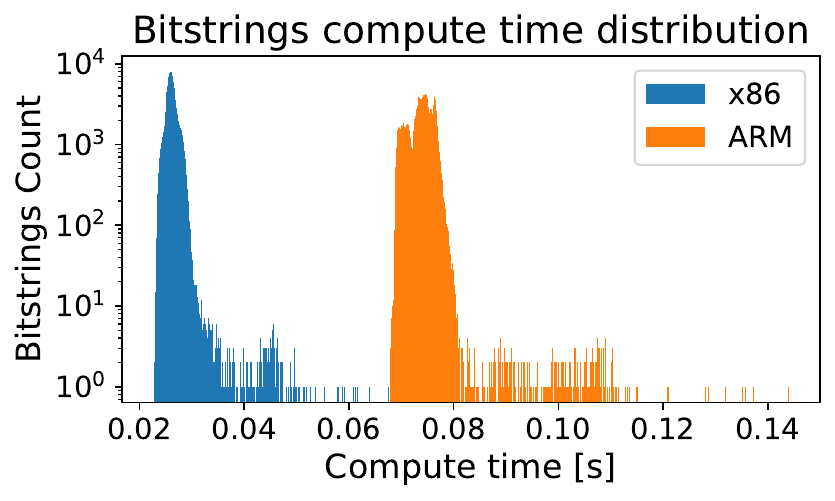}
    \caption{Compute time distribution of $2^{18}$ bitstring pairs for a quantum walk circuit with 18 qubits and 16 steps. Simulation performed with 128 MPI processes, each with 8 OpenMP threads. MPI processes are distributed 16 (6) per node on System A (System B).}
    \label{fig_bitstring_time_dist}
\end{figure}

Figure~\ref{fig_bitstring_time_dist} shows the distribution of kernel compute times (excluding communication) for $2^{18}$ bitstring pairs of an 18-qubit, 16-step quantum walk on these two systems. The runtime varies substantially across outputs because different bitstrings induce different numbers of surviving histories and different amounts of exact pruning. This spread confirms that output costs are highly imbalanced and directly justifies the asynchronous server--worker scheduling used in the distributed implementation.

\begin{figure}[!ht]
    \centering
    \includegraphics[width=\plotwidth]{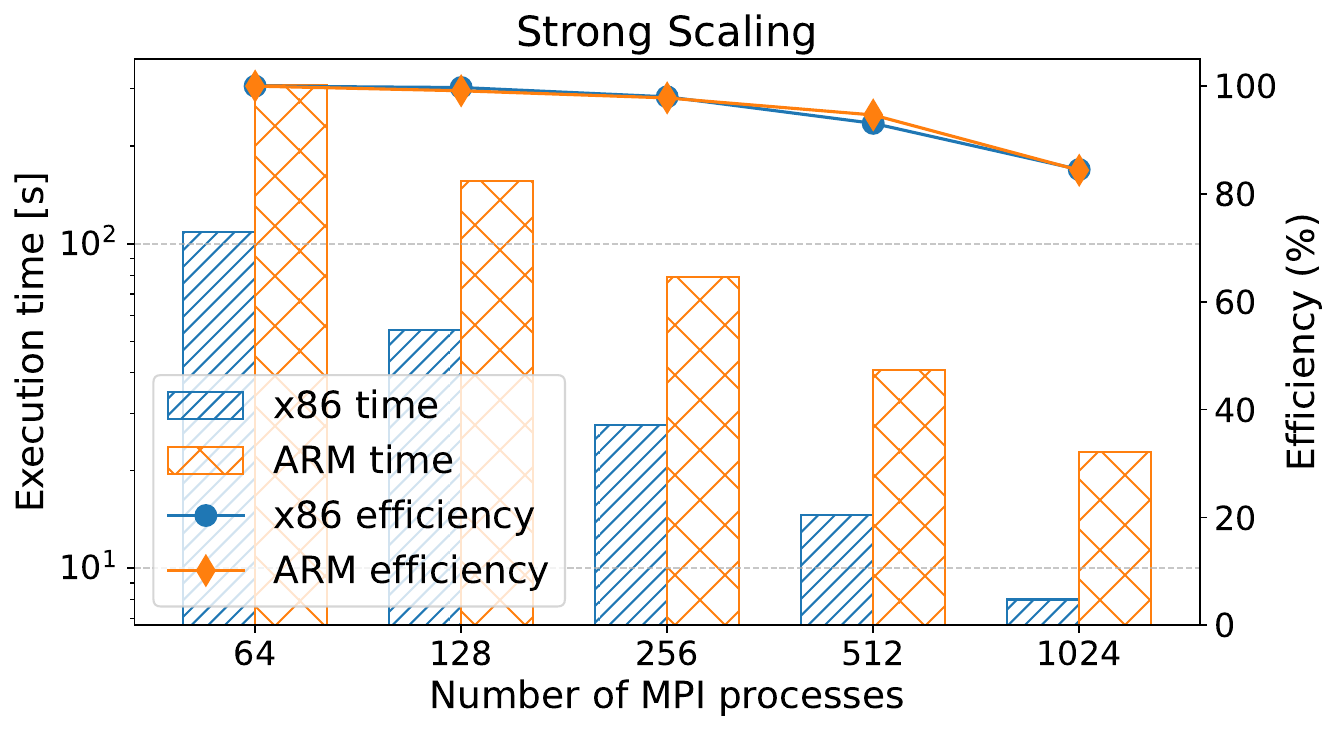}
    \caption{Strong scaling test on AMD EPYC 7742 and Fujitsu A64FX CPUs for a quantum walk circuit with 18 qubits and 16 steps. Each MPI process uses 8 OpenMP threads, and MPI ranks are distributed 16 (6) per node on System A (System B). Efficiency is computed relative to the execution time with 64 MPI processes.}
    \label{fig_strong_scaling}
\end{figure}

Figure~\ref{fig_strong_scaling} reports strong scaling for the same quantum walk circuit from 64 to 1,024 MPI processes. The implementation achieves 93\% parallel efficiency at 512 processes (4,096 cores) and 85\% at 1,024 processes (8,192 cores). 

\begin{figure}[!ht]
    \centering
    \includegraphics[width=\plotwidth]{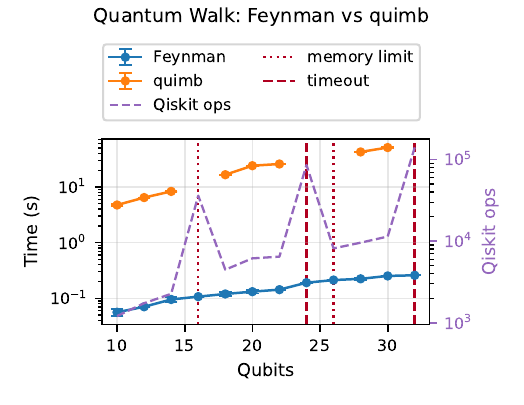}
    \caption{Left axis: Time consumption of our Feynman simulator and quimb. Right axis: Number of transpiled operations used by quimb. Red vertical lines mark qubit counts where quimb exceeded a 10~minute timeout and 64~GiB memory limit.}
    \label{fig_qwalk_quimb}
\end{figure}

Finally, we compare our simulator with the tensor-network simulator quimb. Figure~\ref{fig_qwalk_quimb} shows that our simulator is consistently more than one order of magnitude faster, and that several unfinished quimb runs coincide with peaks in the number of transpiled operations. Part of this advantage comes from our ability to simulate large deterministic gates at low cost.

\section{Discussion and Conclusion}
We presented a distributed method for exact sparse-output quantum circuit simulation based on the pure Feynman sum-over-histories formulation. Instead of evolving the full state vector, the method computes only requested output amplitudes and reduces the effective history space through determinism propagation, artificial sources, checkpointing, and pruning.

The method is most effective when many, but not all, output amplitudes are required and when circuit structure keeps the number of artificial sources small. In this regime it offers an alternative to full-state methods by trading full-state memory growth for output selectivity and structural reduction; it is less favorable for highly scrambling circuits such as random circuit sampling.

Even after reduction, the cost still grows exponentially with the number of artificial sources, making source placement central to performance. The current implementation also does not include noise modeling: stochastic unitary error models could be handled by sampling circuit instances and averaging repeated sparse-output simulations~\cite{isakov2021simulationsquantumcircuitsapproximate, Li_2022}, whereas more general non-unitary channels are harder to integrate in the pure Feynman representation~\cite{PhysRevA.69.062317}. Another direction is a cloud-oriented implementation based on the same output-level decomposition and asynchronous worker model.

\section*{Acknowledgments}
This work was supported by the Swedish Wallenberg Centre for Quantum Technology (WACQT). The authors acknowledge NAISS and EuroHPC for computational resources.



\bibliographystyle{IEEEtran}
\bibliography{main}
\end{document}